\documentclass[twocolumn]{aa}
\usepackage{graphicx}
\usepackage{natbib}
\bibpunct{(}{)}{;}{a}{}{,}

\newcommand{\he}[1] {He\,{\sc #1}}
\newcommand{\hel}[2] {He\,{\sc #1}~$\lambda$#2}

\begin{document}
\title{Time-resolved photometry of the nova remnants \object{DM Gem}, \object{CP Lac}, \object{GI Mon}, \object{V400 Per}, \object{CT Ser} and \object{XX Tau}\thanks{Based in part on observations made with the Jacobus Kapteyn Telescope, which {\em was} operated on the island of La Palma by the Isaac Newton Group in the Spanish Observatorio del Roque de los Muchachos of the Instituto de Astrof\'\i sica de Canarias (IAC), and on observations made with the IAC80 telescope, operated on the island of Tenerife by the IAC in the Spanish Observatorio del Teide of the IAC. Observations were also obtained at the FLWO Observatory, a facility of the Smithsonian Institution.}}

\author{P. Rodr\'\i guez-Gil\inst{1,2} \and
        M. A. P. Torres\inst{3,4}
	}

\offprints{P. Rodr\'\i guez-Gil,\\\email{Pablo.Rodriguez-Gil@warwick.ac.uk}}

\institute{Department of Physics, University of Warwick, Coventry CV4 7AL, UK
\and
Instituto de Astrof\'\i sica de Canarias, V\'\i a L\'actea, s/n, La Laguna, E-38205, Santa Cruz de Tenerife, Spain
\and
Harvard-Smithsonian Center for Astrophysics, 60 Garden St, Cambridge, MA 02138, USA
\and
Physics Department, University College, Cork, Ireland
}

   \date{Received 2004; accepted 2004}

   \abstract{We present the first results of a photometric survey of poorly studied nova remnants in the Northern Hemisphere. The main results are as follows: \object{DM Gem} shows a modulation at 0.123 d (probably linked to the orbit) and rapid variations at $\sim 22$ min. A moderate resolution spectrum taken at the time of the photometric observations shows intense \hel{ii}{4686} and Bowen emission, characteristic of an intermediate polar or a \object{SW Sex} star. Variability at 0.127 d and intense flickering (or quasi-periodic oscillations) are the main features of the light curve of \object{CP Lac}. A 0.1-mag dip lasting for $\sim 45$ min is observed in \object{GI Mon}, which could be an eclipse. A clear modulation (probably related to the orbital motion) either at 0.179 d or 0.152 d is observed in the $B$-band light curve of \object{V400 Per}. The results for \object{CT Ser} point to an orbital period close to 0.16 d. Intense flickering is also characteristic of this old nova. Finally, \object{XX Tau} shows a possible periodic signal near 0.14 d and displays fast variability at $\sim 24$ min. Its brightness seems to be modulated at $\sim 5$ d. We relate this long periodicity to the motion of an eccentric/tilted accretion disc in the binary.
   
   }
   \titlerunning{Time-resolved photometry of nova remnants}
   \authorrunning{P. Rodr\'\i guez-Gil and M. A. P. Torres}
   \maketitle
%

\section{Introduction}
Cataclysmic variable stars (CVs) exhibit their most extreme outbursting behaviour in classical novae. These explosive events are the consequence of thermonuclear runaways in the hydrogen envelope of the white dwarf as a result of continuous mass accretion on to its surface \citep{starrfieldetal76-1}. It takes decades for a nova to return to its quiescent (prenova/postnova) state, where most of them behave like nova-like CVs. Comprehensive reviews on classical novae in particular and CVs in general can be found in \cite{bode+evans89-1} and \cite{warner95-1}, respectively.

So far there is a large number of nova remnants that have received little or no attention at all due to their faintness. Because of the lack of study, a crucial parameter such as the orbital period is currently undetermined for the majority of these quiescent novae. Many of these systems might show the hallmarks of the orbital motion in their optical light curves in the form of pure orbital modulations, superhumps, ellipsoidal modulations, eclipses, etc. Thus, time-resolved photometric monitoring provides an easy way to obtain a reliable estimate of their orbital periods. 

The main goals of this long-term project are to obtain accurate estimates of the orbital periods of poorly studied novae remnants as the necessary input for efficient time-resolved spectroscopic monitoring, and to search for photometric behaviour characteristic of the different CV sub-classes. In this paper we present the first results for six novae remnants.

\section{Observations and data reduction}

\begin{table}[t]
\caption[]{\label{table_obslog}Log of Observations}
\setlength{\tabcolsep}{1.1ex}
\begin{flushleft}
\begin{tabular}{lccc}
\hline
\hline\noalign{\smallskip}
UT date & Coverage & Filter/Grating & Exp.\\   
 & (h) & & (s)\\   
\hline\noalign{\smallskip}
\multicolumn{4}{c}{\textbf{\object{DM Gem}}} \\
\noalign{\smallskip}
\multicolumn{4}{l}{\textbf{1.2-m FLWO CCD photometry}} \\
2004 Mar 11    & 3.98   & White light & 30\\
2004 Mar 14    & 1.28   & White light & 30\\
2004 Mar 15    & 4.87   & White light & 30\\
2004 Mar 16    & 4.87   & White light & 30\\
\noalign{\smallskip}
\multicolumn{4}{l}{\textbf{1.5-m FLWO, FAST spectroscopy}} \\
2004 Mar 15    & ---   & 300 l/mm & 1200\\
\hline\noalign{\smallskip}
\multicolumn{4}{c}{\textbf{\object{CP Lac}}} \\
\noalign{\smallskip}
\multicolumn{4}{l}{\textbf{1.0-m JKT CCD photometry}} \\
2002 Oct 28    & 4.80   & White light & 10\\
2002 Oct 29    & 6.72   & White light & 10\\
2002 Oct 30    & 6.85   & White light & 10\\
2002 Oct 31    & 5.73   & White light & 10\\
\hline\noalign{\smallskip}
\multicolumn{4}{c}{\textbf{\object{GI Mon}}} \\
\noalign{\smallskip}
\multicolumn{4}{l}{\textbf{1.0-m JKT CCD photometry}} \\
2002 Nov 02    & 3.62   & White light & 15\\
2002 Nov 04    & 3.88   & White light & 15\\
\noalign{\smallskip}
\multicolumn{4}{l}{\textbf{1.2-m FLWO CCD photometry}} \\
2004 Mar 17    & 3.84   & White light & 20\\
2004 Mar 18    & 4.04   & White light & 20\\
\hline\noalign{\smallskip}
\multicolumn{4}{c}{\textbf{\object{V400 Per}}} \\
\noalign{\smallskip}
\multicolumn{4}{l}{\textbf{1.0-m JKT CCD photometry}} \\
2002 Nov 03    & 1.57   & $B$ & 360\\
2002 Nov 04    & 6.03   & $B$ & 360\\
\noalign{\smallskip}
\multicolumn{4}{l}{\textbf{0.82-m IAC80 CCD photometry}} \\
2003 Jan 03    & 4.94   & $B$ & 1200\\
\hline\noalign{\smallskip}
\multicolumn{4}{c}{\textbf{\object{CT Ser}}} \\
\noalign{\smallskip}
\multicolumn{4}{l}{\textbf{1.2-m FLWO CCD photometry}} \\
2004 Mar 09    & 3.18   & White light & 20\\
2004 Mar 10    & 2.17   & White light & 20\\
2004 Mar 14    & 3.02   & White light & 20\\
2004 Mar 15    & 3.70   & White light & 20\\
2004 Mar 16    & 4.50   & White light & 20\\
\hline\noalign{\smallskip}
\multicolumn{4}{c}{\textbf{\object{XX Tau}}} \\
\noalign{\smallskip}
\multicolumn{4}{l}{\textbf{1.0-m JKT CCD photometry}} \\
2002 Oct 28    & 4.90   & White light & 240\\
2002 Oct 29    & 4.20   & White light & 150\\
2002 Oct 30    & 4.08   & White light & 180\\
2002 Oct 31    & 3.75   & White light & 150\\
2002 Nov 02    & 6.77   & White light & 150\\
\noalign{\smallskip}\hline
\end{tabular}
\end{flushleft}
\end{table}

Four telescopes were used to obtain the data presented in this paper: the 1.0-m Jacobus Kapteyn Telescope (JKT) on La Palma, the 0.82-m IAC80 telescope on Tenerife, and the 1.2-m and 1.5-m telescopes at the Fred Lawrence Whipple Observatory (FLWO) in Arizona. A log of observations is presented in Table~\ref{table_obslog}, where the novae are listed alphabetically by constellation name. Details on the instrumental setups employed during the observations are given in what follows:

\newcounter{fistro}
\begin{list}{\arabic{fistro}.}{\usecounter{fistro}}
\item The photometric data obtained with the JKT were extracted from CCD images collected without filter (white light) using the 2048$\times$2048 pixel$^2$ SITe CCD camera, with the exception of \object{V400 Per} for which a Johnson $B$ filter was in place.
\item Additional photometry of \object{V400 Per} was obtained with the IAC80 telescope. The telescope was equipped with the 1024$\times$1024 pixel$^2$ Thomson CCD camera and a Johnson $B$ filter.
\item The 1.2-m telescope at FLWO was equipped with the 4-Shooter mosaic CCD camera which consists of an array of four 2048$\times$2048 pixel$^2$ chips. Only the CCD \#3 was used to image the targets. All the images were obtained without filter. 
\item A 20-min optical spectrum of \object{DM Gem} was obtained using the FAST spectrograph \citep{fabricantetal98-1} attached to the 1.5-m telescope at the FLWO. Use of a 3\arcsec~slit and the 300 line mm$^{-1}$ grating provided a wavelength coverage of $\lambda\lambda$3800--7500 with a resolution of $\sim 7$~\AA~(FWHM).  
\end{list}

In order to improve the time resolution of our photometry, the detectors were operated in 2$\times$2 binning mode and only a small window containing the target and a suitable number of comparison stars was read out. All the reduction processes and photometry extraction were done with \texttt{IRAF}\footnote{{\sc iraf} is distributed by the National Optical Astronomy Observatories, which is operated by the Association of Universities for Research in Astronomy, Inc., under contract with the National Science Foundation.}. After correcting the raw images for the effects of bias level and flat field structure, the instrumental magnitudes of each target and three comparison stars were computed by means of aperture photometry. An aperture radius of 1.5 times the full width at half maximum (FWHM) of the typical seeing profile was used \citep{naylor98-1}. The sky background level was measured inside an 8-pixel width annulus with internal radius safely away from the target aperture. The scatter in the comparison light curves shows that the differential photometry is accurate to $< 5$ per cent in all cases.

The spectrum of \object{DM Gem} was extracted and wavelength calibrated
using standard \texttt{IRAF} tasks as part of the CfA spectroscopic data pipeline
\citep{kurtz+mink98-1}. Flux standards taken on the same night were used
to correct for the instrumental response, but the night was not
photometric for an absolute flux calibration.

%
%
%
\section{Results}

\subsection{\object{DM Gem}}

\begin{figure}
\centerline{\includegraphics[width=10cm]{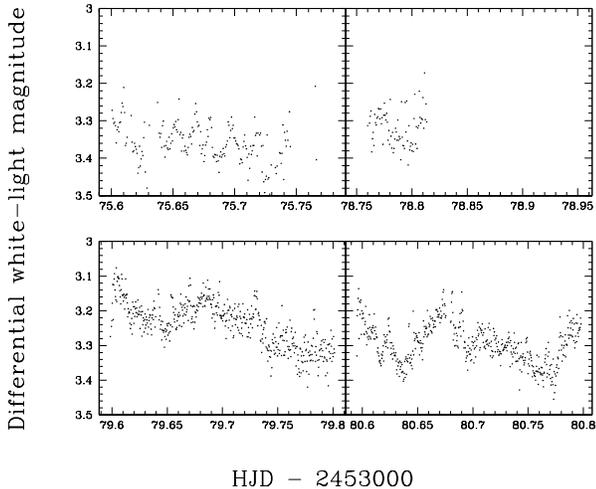}}
\caption[]{\label{fig_dmgem_allcurves} Individual light curves of \object{DM Gem} obtained at the FLWO. From {\em left} to {\em right} and from {\em top} to {\em bottom}: 2004 March 11, 14, 15 and 16.}
\end{figure}

\begin{figure}
\centerline{\includegraphics[width=9cm]{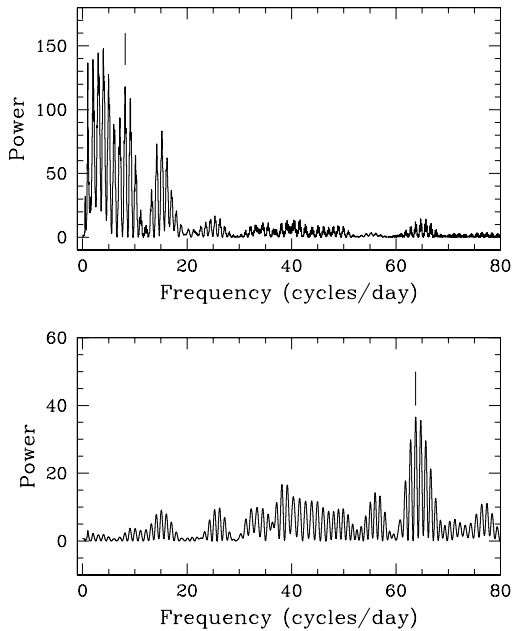}}
\caption[]{\label{fig_dmgem_scargle_foldedP} {\em Top}: Scargle periodogram for \object{DM Gem}. {\em Bottom}: Scargle periodogram after removing the 0.123-d signal.}
\end{figure}

\object{DM Gem} (Nova Geminorum 1903) was discovered at m${_\mathrm{pg}}\sim8.0$ on 1903 March 16 by Professor Turner. Examination of archival
photographic plates showed the nova at m${_\mathrm{pg}}\sim5$ on 1903 March 6
\citep{pickeringetal03-1}. A $t_2=6$ days \citep{schmidt57-01} is
consistent with a very fast light curve, where $t_2$ (or $t_3$) is 
the time in days that the light curve takes to decline 2 (or 3) magnitudes 
below maximum brightness. A magnitude of $V=17.38$ for the postnova was reported by \cite{szkody94-1}.

We observed \object{DM Gem} on the nights of 2004 March 11, 14, 15 and 16 at the FLWO. The light curves (Fig.~\ref{fig_dmgem_allcurves}) seem to show a non-sinusoidal variation at about 3 hours (see e.g. the 2004 March 16 light curve). We have combined all the data to compute the Scargle periodogram \citep{scargle82-1} presented in the upper panel of Fig.~\ref{fig_dmgem_scargle_foldedP}. The strongest peak near 3 hours is centred at $0.123 \pm 0.002$ d. Hereafter the quoted errors represent half the FWHM of the corresponding peak as measured by using a Gaussian fit. The peaks at lower frequencies are centred at multiples of 1 d$^{-1}$ as a consequence of the observing window. The peak at 0.123 d is close to this region of false frequencies. However, the fact that we see a modulation at this period in two of the light curves indicates that it might be real. Note that \cite{lipkinetal00-1} found the same signal, which was interpreted as a quasi-periodic oscillation. 

The light curve of \object{DM Gem} also displays short-term variations at a time scale of $\sim 20$ min (see e.g. the 2004 March 11 light curve). In order to search for this periodicity, a new Scargle periodogram was computed after removing the 0.123-d signal from the individual light curves by subtracting a smoothed version of each light curve (bottom panel of Fig.~\ref{fig_dmgem_scargle_foldedP}). The strongest peak is centred at $22.58 \pm 0.09$ min.

\begin{figure}
\centerline{\includegraphics[width=9cm]{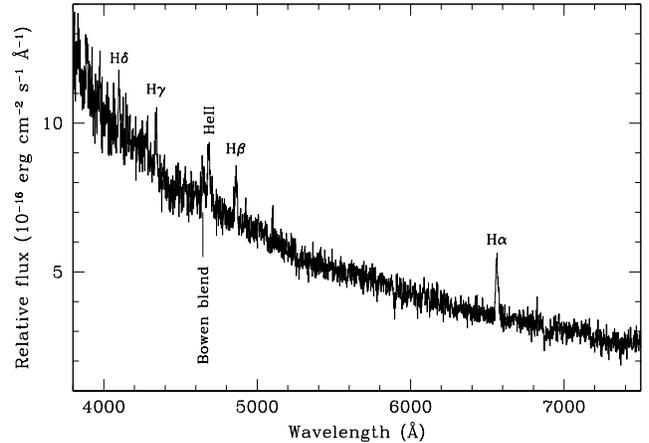}}
\caption[]{\label{fig_dmgem_spectrum} Optical spectrum of \object{DM Gem}.}
\end{figure}

A single spectrum of \object{DM Gem} was obtained on the night of 2004 March 15. The spectrum (Fig.~\ref{fig_dmgem_spectrum}) shows single-peaked emission lines of \hel{ii}{4686} (equivalent width, EW=$5\pm1$~\AA,
FWHM=$1088\pm130$~km s$^{-1}$), the Bowen blend at $\lambda\lambda4640-4650$ (EW=$3\pm1$~\AA, FWHM=$1100\pm260$~km s$^{-1}$) and the
Balmer series up to H$\delta$: H$\alpha$ (EW=$11\pm1$~\AA, FWHM=$914\pm50$~km
s$^{-1}$), H$\beta$ (EW=$4\pm1$~\AA, FWHM=$1110\pm120$~km s$^{-1}$), H$\gamma$ (EW=$3\pm1$~\AA, FWHM=$900\pm140$~km s$^{-1}$) and H$\delta$ (EW=$2\pm1$~\AA, FWHM=$800\pm140$~km s$^{-1}$). The unresolved NaD doublet at $\lambda5890$, $\lambda5896$ is also evident with an EW of $-1.5\pm1$~\AA. This large EW prevents us from using the \cite{munari+zwitter97-1} calibration to derive the reddening from this interstellar feature as the NaD EW/reddening relationship does not change appreciably for EW $\leq -0.5$~\AA. A power-law fit to the observed spectrum in the range $\lambda\lambda4000-6900$ (after masking the emission lines) provides a power-law index of $-2.19\pm0.03$.

The most remarkable spectral feature in \object{DM Gem} is the strong \ion{He}{ii} $\lambda$4686 emission line (\ion{He}{ii}/H$\beta$ $\ga 1$), which indicates the presence of a source of ionising photons within the binary. This, together with the (unconfirmed) rapid photometric variability, are characteristic signatures of the Intermediate Polar (IP) CVs and many \object{SW Sex} stars. The latter have been proposed as magnetic accretors (\citealt{rodriguez-giletal01-1,rodriguez-gil03-1}). Intense \hel{ii}{4686} emission has been observed e.g. in the IP systems \object{V1223 Sgr} \citep{steineretal81-1} and \object{AO Psc} \citep{cordovaetal83-1}, and in the \object{SW Sex} stars \object{BT Mon} \citep{williams83-1} and \object{V348 Pup} \citep{rodriguez-giletal01-2}.

\subsection{\object{CP Lac}}

\object{CP Lac} (Nova Lacertae 1936) was discovered visually by Kazuaki Gomi
during the 1936 June 18 total solar eclipse. The nova reached maximum
light ($V\simeq2.0$; \citealt{bertaud45-1}) two days later, being
$\sim13.5$~mag brighter than the pre/postnova \citep{robinson75-1}.
The light curve during the decline phase after the nova explosion
showed that \object{CP Lac} was a very fast nova with t${_2}=5.3$~days, t${_3}=
9.8$~days \citep{bertaud45-1}. 

The postnova optical spectrum \citep{ringwaldetal96-1} shows Balmer and \he{i} emission lines. Also present are the \hel{ii}{4686} emission line and the Bowen blend, both with similar strength. A multi-year (1991--1997) light curve \citep{honeycuttetal98-2} shows six stunted outbursts characterised by having smaller amplitude than typical dwarf novae outbursts.

\begin{figure}
\centerline{\includegraphics[width=9cm]{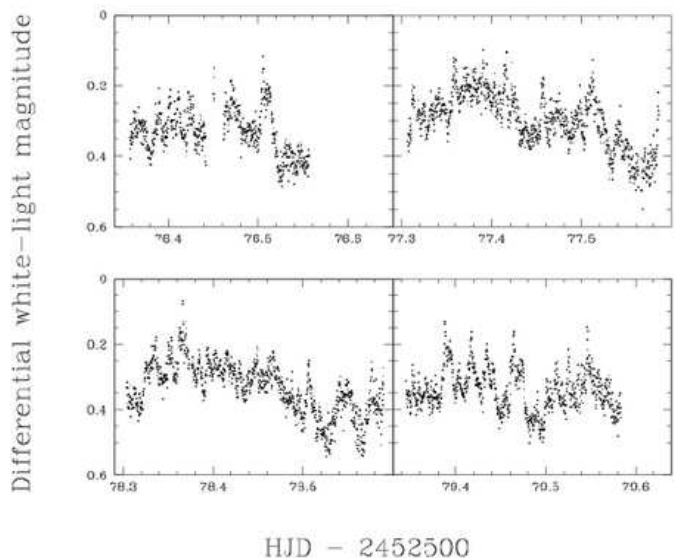}}
\caption[]{\label{fig_cplac_allcurves} Unfiltered JKT light curves of \object{CP Lac}. {\em Top panel}: 2002 October 28 ({\em left}) and 29 ({\em right}). {\em Bottom panel}: 2002 October 30 ({\em left}) and 31 ({\em right}).}
\end{figure}

\begin{figure}
\centerline{\includegraphics[width=9cm]{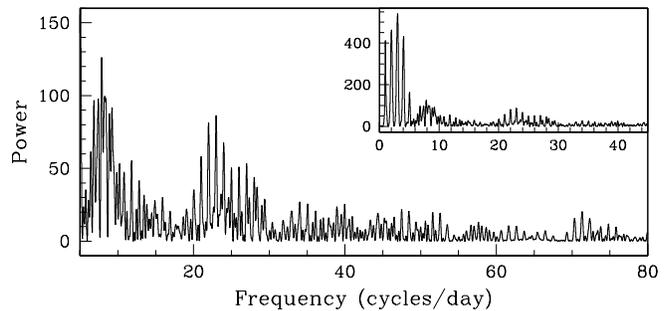}}
\caption[]{\label{fig_cplac_scargle} Scargle periodogram computed from all the \object{CP Lac} light curves (nightly averages subtracted). The strong peaks in the low frequency regime of the periodogram ($\nu < 5$ d$^{-1}$) are very likely due to the observing window.}
\end{figure}

In Fig.~\ref{fig_cplac_allcurves} we present the unfiltered light
curves of \object{CP Lac} obtained on 2002 October 28-31 with the JKT
on La Palma. The light curves are dominated by intense and rapid variations (tens of minutes time scale) with an
amplitude of $\sim 0.2$ mag. A possible variation at $\sim 0.1$
d superimposed on a decreasing longer-term trend might be also
present (see e.g. the 2002 October 29 light curve). A combined linear
plus sine fit to the October 29 light curve gives a period of $\sim
0.12$ d and a semi-amplitude of $\sim 0.05$ mag. To search for this
and other periodicities in the light curves, we computed the Scargle
periodogram after subtracting the nightly average values. The
resulting power spectrum shows a group of strong peaks at low
frequencies centred at approximately integer values and separated by 1
d$^{-1}$ (see inset plot in Fig.~\ref{fig_cplac_scargle}). They are
the result of the observing window. Two groups of peaks are located
around $8$ d$^{-1}$ and $23$ d$^{-1}$. The centre of the strongest
peak in each group was measured by fitting a Gaussian function to
yield $0.127 \pm 0.002$ d (in agreement with our initial estimate) and
$0.0435 \pm 0.0002$ d ($62.7 \pm 0.3$ min).

\begin{figure}
\centerline{\includegraphics[width=7cm]{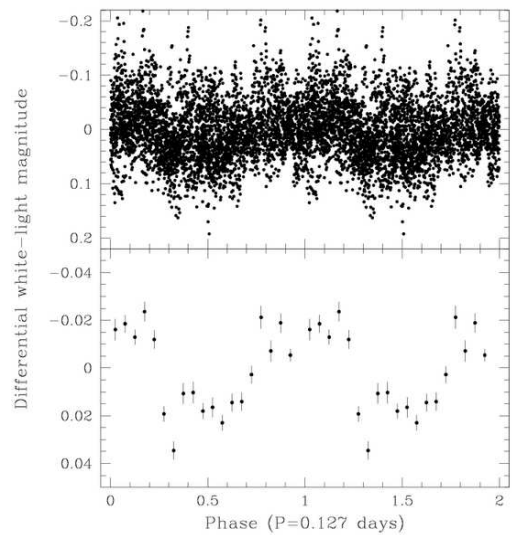}}
\caption[]{\label{fig_cplac_folded} \object{CP Lac} light curves folded on the 0.127-d period after subtracting the long term trend (see text for details). The bottom panel shows the folded light curve after binning the data points into 20 phase intervals. A full cycle has been plotted twice for continuity.}
\end{figure}

We detrended the light curves of the longer term variation on which
the 0.127-d wave lies. To do this, a smoothed version of each light
curve (using a 300-point boxcar) was subtracted to the original
ones. We folded the detrended light curves on the 0.127-d period,
obtaining the curve shown in Fig.~\ref{fig_cplac_folded}. The time of zero phase corresponds to the
first acquired data point. The binned light curve shows a narrow dip at relative phase $\sim 0.3$ lasting for
$\sim 25$ min. It would be desirable to have a longer photometric coverage to check whether this dip is an eclipse or not. Such a shallow eclipse (if present at all) can  be lost in the similar amplitude, short-term variations observed in the light curve.

If the variation at 0.127 d is related to the orbital motion, \object{CP Lac} will lie just above the period gap, a region mainly inhabited by the \object{SW Sex} stars. The strength of the \hel{ii}{4686} emission line and the rapid and non-coherent photometric variability in \object{CP Lac} also relate it to this class of CVs. Time-resolved spectroscopy of this fairly bright object will definitely shed more light on this issue.  

\subsection{\object{GI Mon}}

\begin{figure}
\centerline{\includegraphics[width=10cm]{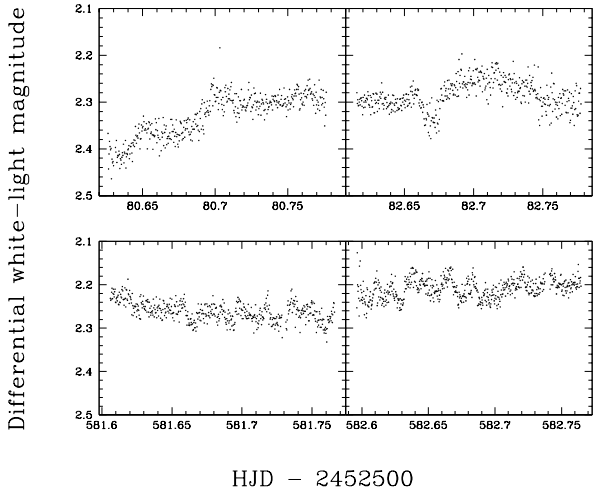}}
\caption[]{\label{fig_gimon_allcurves} Unfiltered light curves of \object{GI mon} obtained with the JKT ({\em top panels}; 2002 November 2, 4) and at the FLWO ({\em bottom panels}; 2004 March 17--18).}
\end{figure}

\object{GI Mon} (Nova Monocerotis 1918) was a fast nova discovered on 1918
February 4 by M. Wolf after maximum brightness at m${_\mathrm{pg}}=8.5$ (see
e.g. \citealt{pickering18-1}). The postnova has a magnitude of $V=16.2$
\citep{szkody94-1} and a blue spectrum with weak Balmer and \hel{ii}{4686} emission lines (see \citealt{liu+hu00-1} and references
therein). No nova shell was evident in the postnova spectrum at
$\sim 6555$~\AA~\citep{gill+obrien98-1}.

We collected data on \object{GI Mon} during 2002 November 1 and 4 (JKT) and 2004 March 17--18 (FLWO). The light curves are presented in Fig.~\ref{fig_gimon_allcurves}. The 2002 November 4 light curve displays a $\sim 0.1$-mag, $\sim 45$-min long dip that could be interpreted as an eclipse. If this is the case, the absence of a similar dip in the other light curves would support a long orbital period. Additionally, the 2004 light curves show low amplitude, short time scale ($\sim 10$ min) variations which are not present in the 2002 data.

One day after collecting our last data on \object{GI Mon} we were aware of the results of \cite{woudtetal04-1}, who claim an orbital or superhump period of 0.1802 d. Our periodogram from the 2002 dataset shows a cluster of peaks centred at approximately this period. However, there is no sign of it in the periodogram computed using our 2004 dataset despite the fact that the nightly observing windows are roughly the same in both years. Finally, the light curve labelled ``S6155'' in figure 9 of \cite{woudtetal04-1} shows a $\sim 0.1$-mag dip with a duration of $\sim 40$ min, which is similar to the dip we found in our data.

\subsection{\object{V400 Per}}

\object{V400 Per} (Nova Persei 1974) was discovered spectroscopically after
maximum at m${_\mathrm{pg}}\sim11$ on 1974 November 9 by
\cite{sanduleak74-1}. \cite{shao74-1} and \cite{scovil74-1}
independently identified the prenova with a faint blue star with
m${_\mathrm{pg}}\sim19.5$. Examination of prediscovery photographic plates
showed that the nova reached $V\sim8$ on September 24 and there was no
evidence for earlier outbursts during the period $1898.8-1952.8$
(\citealt{shao74-1}; \citealt{liller74-1}). The nova light curve of
\object{V400 Per} was in the limit of a moderately fast to slow nova
(\citealt{mattei75-1}; \citealt{landolt75-1}). A $V$ magnitude of 19 for
the postnova was reported in \cite{szkody94-1}. 

\begin{figure}
\centerline{\includegraphics[width=9cm]{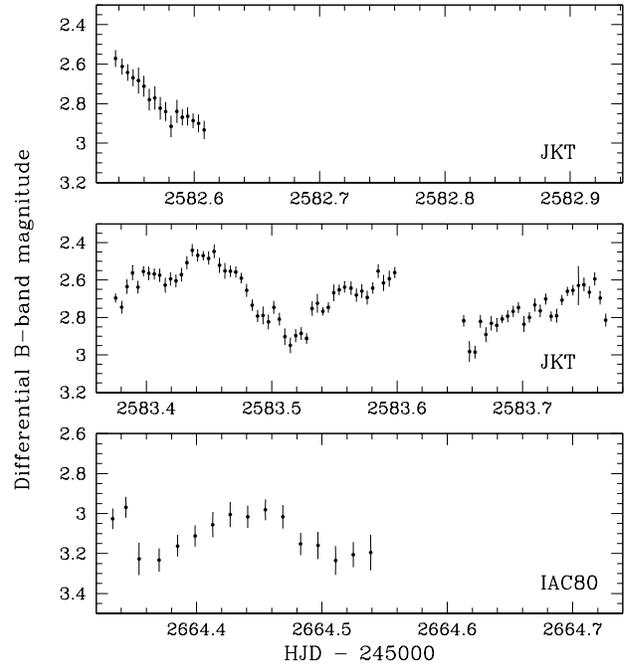}}
\caption[]{\label{fig-v400per-rawlcs} $B$-band light curves of \object{V400 Per}. From {\em top} to {\em bottom}: 2002 November 3--4 (JKT), and 2003 January 3 (IAC80). A modulation with a period of $\sim 0.16$ d can be seen in the curve with longest coverage.}
\end{figure}

It proved impossible to take useful images of \object{V400 Per} in white light and $2 \times 2$ binning due to the proximity (24.5\arcsec~separation) of a bright star with $R \approx 11.0$ (USNO 1350 03079208). We then decided to use the Johnson $B$ filter to minimise its contribution, with the disadvantage of having to increase the exposure times. The CCD window was also chosen as to exclude as much of the contaminating star from the frames as possible.

\begin{figure}
\centerline{\includegraphics[width=9cm]{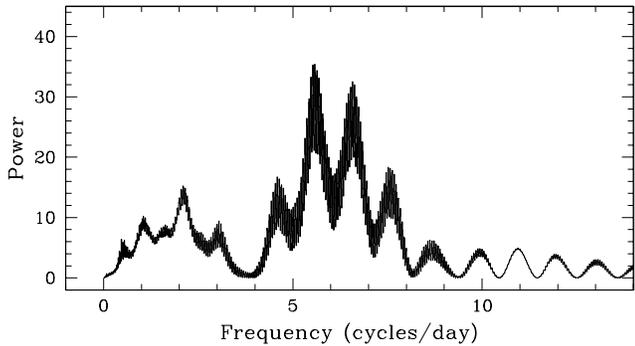}}
\caption[]{\label{fig-v400per-scargle} Scargle periodogram computed from all the photometric data on \object{V400 Per} after subtracting the nightly averages.}
\end{figure}

The differential $B$-band light curves obtained on 2002 November 3 and 4 (JKT), and on 2003 January 3 (IAC80) are presented in Fig.~\ref{fig-v400per-rawlcs}. The curve with longer coverage shows a modulation with a likely period of $\sim 0.16$ d and a semi-amplitude of $\sim 0.2$ mag. A sine fit to the IAC80 light curve provided a period of $0.16 \pm 0.02$ d with a semi-amplitude of $0.12 \pm 0.2$ mag. A Scargle periodogram computed from all the data (after subtracting the nightly means) is shown in Fig.~\ref{fig-v400per-scargle}. Two prominent peaks are seen in the frequency interval $5-7$ d$^{-1}$, centred at $5.58 \pm 0.03$ and $6.58 \pm 0.03$  d$^{-1}$ (a 1 d$^{-1}$ alias). These frequencies correspond to the periods $0.179 \pm 0.001$ d and $0.152 \pm 0.001$ d, respectively. Unfortunately, the amount of data available is insufficient to provide an unambiguous period. 
The observed modulation is likely related to the binary orbital motion, either as an orbital modulation or a superhump.

\begin{figure}
\centerline{\includegraphics[width=10cm]{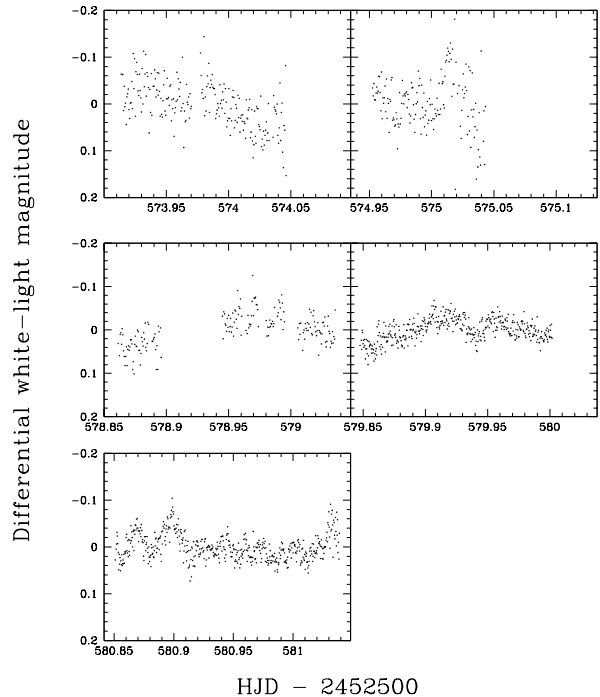}}
\caption[]{\label{fig_ctser_allcurves} Unfiltered light curves of \object{CT Ser} taken at the FLWO. From {\em left} to {\em right} and from {\em top} to {\em bottom}: 2004 March 9, 10, 14, 15 and 16.}
\end{figure}

\subsection{\object{CT Ser}}
\object{CT Ser} (Nova Serpentis 1948) was likely a slow nova discovered after
maximum brightness at magnitude close to 9.0 by Bartay on 1948 April 9 (see \citealt{mclaughlin48-1}). From
the analysis of a 2-hour postnova light curve, \cite{howelletal90-1}
reported photometric periodicites at 13 and 65 min, suggesting an
orbital origin to the latter. Additionally, \cite{hoardetal02-2}
found a significative change in the near-IR colours of \object{CT Ser} with respect to those reported in \cite{harrison92-1} from observations obtained
six years earlier. This is explained as due to a transient phenomenon
rather than to colour evolution. The postnova optical
spectrum of \object{CT Ser} shows a blue continuum with Balmer and \hel{ii}{4686} lines in emission \citep{ringwaldetal96-1,szkody+howell92-1}. However, not all of the H$\alpha$ emission has its origin in the binary system as shown by \cite{downes+duerbeck00-1} who resolved the nova shell at this
wavelength.

\begin{figure}
\centerline{\includegraphics[width=9cm]{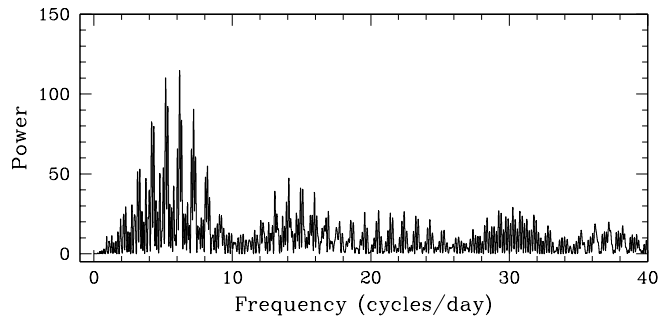}}
\caption[]{\label{fig_ctser_scargle} Scargle periodogram for \object{CT Ser} after subtracting the nightly means.}
\end{figure}

The light curves of \object{CT Ser} obtained on 2004 March 9--10 and 14--16 at the FLWO are shown in Fig.~\ref{fig_ctser_allcurves}. The differential photometry was made with respect to different comparison stars so we have subtracted the nightly averages. 
The Scargle periodogram obtained from our data is shown in Fig.~\ref{fig_ctser_scargle}. The strongest peak is centred at $0.162 \pm 0.001$ d. Alas, we can not address any conclusion since our longest light curve only spans $\sim 0.19$ d and the data are not widely spread in time. 
\object{CT Ser} also shows shorter time scale variability, but no clear peak is observed in the periodogram. Thus, these variations are likely due to erratic flickering.  

\subsection{\object{XX Tau}}

\object{XX Tau} (Nova Tauri 1927) was discovered by \cite{schwassmann+wachmann28-1}.  The light curve of this fast nova attained photographic
magnitude m${_\mathrm{pg}}=6$ and displayed t${_2}=24$~days, t${_3}=43$~days
(see e.g.  \citealt{downes+duerbeck00-1}).  The postnova has an
uncertain visual magnitude of $V=19.8$ \citep{szkody94-1}.

\begin{figure}
\centerline{\includegraphics[width=9cm]{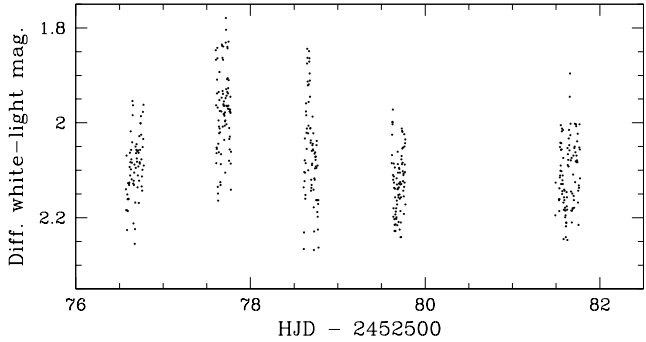}}
\caption[]{\label{fig_xxtau_longterm} Long-term light curve of \object{XX Tau} in white light. The plot shows a clear variation of the mean brightness from night to night.}
\end{figure}

\begin{figure}
\centerline{\includegraphics[width=9cm]{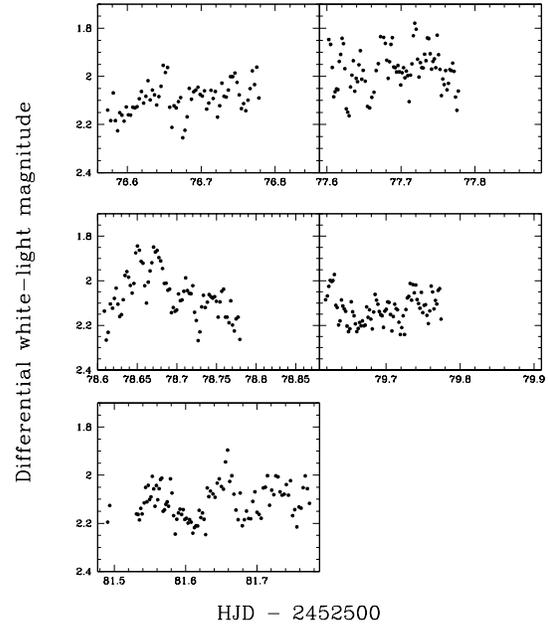}}
\caption[]{\label{fig_xxtau_allcurves} Individual light curves of \object{XX Tau} obtained with the JKT. From {\em left} to {\em right} and from {\em top} to {\em bottom}: 2002 October 28--31 and November 2.}
\end{figure}

\begin{figure}
\centerline{\includegraphics[width=9cm]{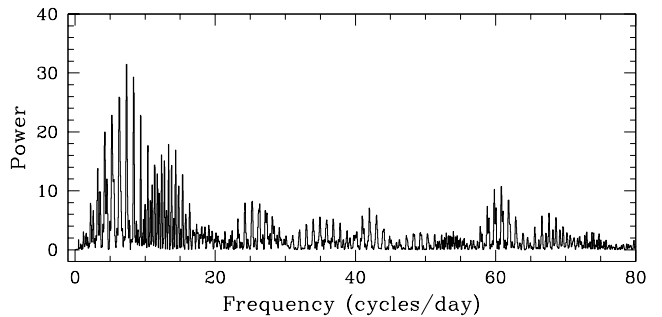}}
\caption[]{\label{fig_xxtau_scargle} Scargle periodogram for \object{XX Tau}.}
\end{figure}

The long term light curve of \object{XX Tau} is shown in Fig.~\ref{fig_xxtau_longterm}. The mean brightness seems to be modulated with a periodicity of the order of a few days that could be associated to a long-term variation in the system such as that due to the precession of an eccentric/tilted accretion disc. No obvious periodic variability is observed in the individual light curves (Fig.~\ref{fig_xxtau_allcurves}). The observed scattering reflects the presence of a short time scale variation with an
amplitude $\la 0.2$ mag. 
We have constructed a Scargle periodogram (Fig.~\ref{fig_xxtau_scargle}) using
all the available data. Most of the power is concentrated at about 7
d$^{-1}$, with the strongest peak centred at $0.136 \pm 0.002$ d.
The periodogram also shows significant power around 60 d$^{-1}$,
peaking at $0.01645 \pm 0.00002$ d ($23.69 \pm 0.03$ min). 
However, folding the data on 23.69 min (after removing the
0.136-d modulation) does not yield a clear modulation and thereby we
can be dealing with flickering or quasi-periodic oscillations rather
than with a coherent signal.

The long term varibility of \object{XX Tau} may have a $\sim 5$-d periodicity. If this is the case, it will be similar to that observed in the \object{SW Sex} stars \object{V442 Oph} and \object{RX J1643.7+3402} \citep{pattersonetal02-1}, and \object{PX And} \citep{stanishevetal02-1}. These systems show a modulation of $4-5$ d attributed to the precession of an eccentric/tilted disc and have orbital periods in the range $\sim 0.12-0.14$ d. Therefore, the modulation at 0.136 d we see in \object{XX Tau} could be related to the orbital period (e.g. a superhump period).

\section{Conclusions}

\begin{figure}
\centerline{\includegraphics[width=9cm]{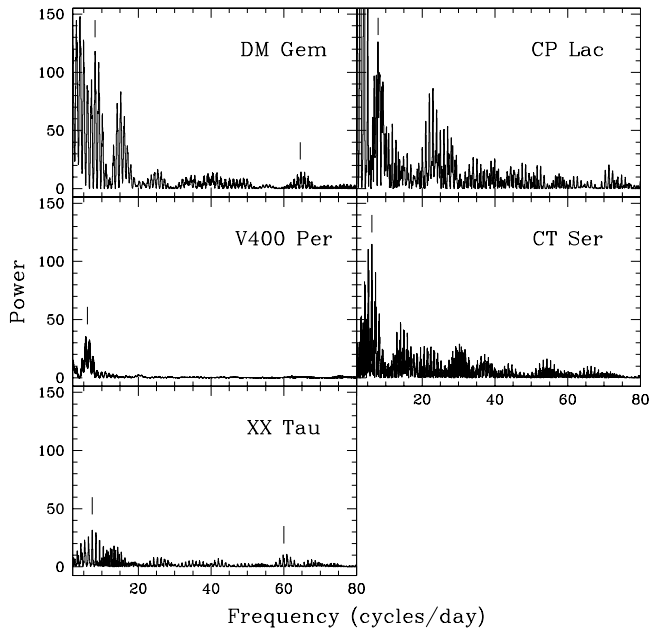}}
\caption[]{\label{fig_allperiodograms} All Scargle periodograms presented in the paper (with the exception of \object{GI Mon}) for easy comparison between objects. The peaks corresponding to the possible periodicities listed in Table~\ref{table_allperiodograms} (excepting the long term variation in \object{XX Tau}) are marked with vertical lines.}
\end{figure}

In Fig.~\ref{fig_allperiodograms} we show all the periodograms presented in the paper (with the exception of \object{GI Mon}) so that the excess power for a given object can be easily spotted by comparison with the others. The likely periodicities found are summarised in Table~\ref{table_allperiodograms}.

\subsection{\object{DM Gem}}  
The light curves provide a modulation at 0.123 d and possible rapid variability at $\sim 22$ min. The continuum slope of the optical spectrum suggests the presence of an accretion disc in the system. The strength of the \hel{ii}{4686} emission line is comparable to that of H$\beta$. The Bowen blend is also intense, indicating the presence of a source of ionising photons. We point out that \object{DM Gem} might be an IP or a \object{SW Sex} star. Future  spectroscopic studies must have the appropriate time resolution to sample a possible $\sim 20$-min modulation in the emission-line flux and enough time coverage to properly deal with a $\sim 3$-hour orbital period.

\subsection{\object{CP Lac}}   
This nova remnant is characterised by intense short time scale variability superimposed on a 0.127-d wave. We tentatively suggest that shallow eclipses might be present. The intense \hel{ii}{4686} and Bowen emission in its optical spectrum points towards a magnetic nature. If the 0.127-d periodicity is related to the orbital motion, \object{CP Lac} will lie within the period range dominated by the \object{SW Sex} stars.

\subsection{\object{GI Mon}}
Our data are insufficient to confirm the 0.1802-d periodicity reported by \cite{woudtetal04-1}. We recorded a shallow dip in one of our light curves which must be confirmed as an eclipse. In the case \object{GI Mon} is an eclipsing CV, the length of our longest light curve constrains the orbital period to be $> 0.2$ d.

\subsection{\object{V400 Per}}
Our $B$-band photometry provides a clear modulation at either 4.30 or 3.65 hours, which we link to the orbital motion of the binary.

\subsection{\object{CT Ser}}
The light curves do not provide a clear picture. A modulation at $\sim 0.17$ d might be present. Erratic, short time scale variability is also observed.

\subsection{\object{XX Tau}}
Our results favour a modulation at $\sim 0.14$ d for this nova remnant. Fast variability at $\sim 24$ min might also be present, but it is uncertain. The most obvious feature is a brightness modulation at $\sim 5$ d, similar to other nova-like CVs (particularly the \object{SW Sex} stars). We attribute this long-period variation to the motion of an eccentric/tilted accretion disc in the binary.

\begin{table}
\caption[]{\label{table_allperiodograms}Summary of possible periodicities}
\begin{flushleft}
\begin{tabular}{lll}
\hline
\hline\noalign{\smallskip}
Object & Periodicity (min) & Remarks\\   
\hline\noalign{\smallskip}
\object{DM Gem}   & 177.1, 22.6 & Strong \hel{ii}{4686}. Magnetic? \\
\object{CP Lac}   & 182.9       & Shallow eclipses? SW Sex? \\
\object{GI Mon}   & 259.2       & \\
\object{V400 Per} & 219 or 258  & \\
\object{CT Ser}   & 233.3       & \\
\object{XX Tau}   & 7200, 195.8, 23.7 & Disc precession? \\
\noalign{\smallskip}\hline
\end{tabular}
\end{flushleft}
\end{table}

\begin{acknowledgements}

We thank the anonymous referee for his/her valuable comments on the original manuscript. We are very grateful to Perry Berlind who
kindly obtained the spectrum of \object{DM Gem} and to Emilio Falco
for allocating two extra nights to our project. PRG thanks PPARC for
support through a PDRA grant. This work was supported in part by NASA
grant NAG5-9930.

\end{acknowledgements}

\bibliographystyle{aa}
\bibliography{aamnem99,$HOME/tex/aabib}
\end{document}